\documentclass[12pt]{article}
\usepackage{graphicx}
\begin{document}

\pagestyle{empty}

\noindent
{\bf Long Secondary Periods in Pulsating Red Giants: A Century of Investigation}

\bigskip

\noindent
{\bf John R. Percy\\Department of Astronomy and Astrophysics, and\\Dunlap Institute of Astronomy and Astrophysics\\University of Toronto\\Toronto ON\\Canada M5S 3H4}

\smallskip

\noindent
{\bf Email: john.percy@utoronto.ca}

\medskip

\bigskip

{\bf Abstract}  Red giants (RGs) are unstable to radial pulsation.  About
a third of them also show a long secondary period (LSP), 5-10 times the
pulsation period.  The LSPs were recently ascribed to eclipses of the RG
by a low-mass dust-enshrouded companion.  LSPs have been known for over a century.  In
this paper, I use primarily AAVSO visual and photoelectric observations
to look for evidence of LSPs in 103 red giant stars listed by Nancy Houk in 1963 as
having LSPs, based mostly on photographic photometry.  I have
determined LSPs in 37 stars, and upper limits (some of them not very
stringent) in 25.  In the former, the
ratio of LSP to pulsation period peaks strongly at 10, which suggests that
most of the stars are pulsating in the first overtone.  The LSPs are
consistent with Houk's in 33 of the 37 stars.   I have identified
16 stars as bimodal pulsators; their period ratios are consistent with previous
observational and theoretical results.  For 14 stars, the periods
in the {\it General Catalogue of Variable Stars} are incorrect or absent.

% \medskip

% \noindent
% AAVSO keywords = photometry, CCD; pulsating variables; giants, red; period analysis; amplitude analysis; globular clusters

\medskip

\noindent
ADS keywords = stars; stars: late-type; techniques: photometric; methods: statistical; stars: variable; stars: oscillations; clusters, globular

\medskip

%\noindent
%{\it Note to referee and editor: instead of including the long appendix of
%notes about individual stars, I could post this in a permanent repository
%which is maintained by my university's library.}

%\medskip

\noindent
{\bf 1. Introduction}

\smallskip

Red giants (RGs) are unstable to pulsation in one or more radial modes.  
In general, the larger and brighter the star, the longer the pulsation 
period (P) and the larger the amplitude (A).  About a third of RGs also have a long
secondary period (LSP), 5 to 10 times longer than the dominant pulsation
period, depending on whether the pulsation period is the fundamental or first
overtone (e.g. Wood 2000).  Until recently, the cause of the LSP was uncertain, but Soszy\'{n}ski
{\it et al}. (2021) have made a strong case for a binary mechanism: the companion was
initially a planet, which later accreted material from the RG wind, and grew into
a brown dwarf or low-mass star, which produces the observed LSP velocity
variations in the RG.  The dust-enshrouded companion eclipses the RG periodically, producing the
LSP.

LSPs have been known for almost a century; a series of three papers helped to
bring them to attention (O'Connell 1933, Payne-Gaposchkin 1954, and Houk 1963).
These papers use photographic photometry, some of it extending back to the late 19th century.  O'Connell (1933) pointed out that the ratio of LSP/P was about 10, but only four of his
stars were actually red giants, and his result depends strongly on one star -- V Hya.  Payne-Gaposchkin (1954) collected information
on many more red giants, of various types, (but even included RR Lyrae stars
showing the Blazhko Effect, which is a totally different phenomenon).  Houk (1963),
in her Table III, lists 103 PRGs with LSPs.

I was interested in how effective the photographic photometry was in identifying and 
measuring LSPs, and whether the LSPs would stand up with time.  Therefore,
in the present paper, I use visual v and photoelectric V observations from
the American Association of Variable Star Observers (AAVSO) International
Database, and V observations from the All-Sky Automated Survey for Supernovae
(ASAS-SN) to investigate the primary and secondary periods, and to compare
them with the periods listed in Houk (1963).  Previous studies of LSPs in RGs have been carried out on selected stars,
using the same databases.  These include Percy and Deibert (2016) and
Percy and Leung (2017), who studied
RGs with long and dense AAVSO data, and Percy and Fenaux (2019), who carried
out a pilot study of RGs using the much-shorter ASAS-SN data.

\medskip

\noindent
{\bf 2. Data and Analysis}

\smallskip

For the 103 stars listed in Table III of Houk (1963), I
analyzed visual (v) and photoelectric (V) data from the AAVSO International
Database (AID), and V data from the ASAS-SN variable star database and catalog
(Shappee {\it et al}. 2014, Jayasinghe {\it et al.} 2018, 2019), using
light-curve analysis
and time-series analysis using the AAVSO VSTAR software package (Benn 2013).
Note that VSTAR provides the semi-amplitude SA, not the full amplitude or range.
The ASAS-SN data and light curves are freely available on-line
(asas-sn.osu.edu/variables) as are the AID data and light curves, and
the VSTAR package (aavso.org).
The AAVSO v observations are limited by their low precision -- typically
0.2 to 0.3 mag -- but are very numerous for many of the stars.  
The AAVSO V observations tend to be sparse.  The ASAS-SN
observations extend over only a few years.

In some cases, a strategy which was helpful for determining the LSP was
to bin the visual observations, in bins equal in length to the pulsation
period.  This strategy was less-useful when the SA was less than 0.1 mag,
or when the star was a bimodal pulsator, or when the data were sparse
and/or had unusually long seasonal or other gaps.

\medskip

\noindent
{\bf 3. Results}

\smallskip

Table 1 lists the 37 stars for which I found good 
evidence for an LSP; Table 2 lists 25 stars for which I found some 
evidence that an LSP is {\it not} present; we include a rough upper limit to
the semi-amplitude that I could detect.  In some cases, the upper limit
was too large to be of much significance.   Table 3 lists 16 stars which may be
bimodal.  Table 4 lists 15 stars for which the period in the {\it General Catalogue
of Variable Stars} (Samus {\it et al}. 2017) is probably not correct, or
is absent.  Note that,
for some stars in the sample, the {\it GCVS} period is the LSP, not the pulsation period.

For the following stars, I obtained no significant results about possible
bimodal pulsation, or about LSPs, but there may be other useful information
in the Notes, in the Appendix: SS And, V Aql, RX Boo, ST Cam, 
RS Cap, BZ Car, $\mu$ Cep, CV Cep, 
T Crt, 
RV Cyg, CH Cyg, V539 Cyg, EI Del,
S Dra, UU Dra, V Eri, SU Hya, U Lac, WY Lac, KP Lyr, Y Mic, RV Mon, SX Mon, 
W Nor, TW Oph, 
V521 Oph, V574 Oph, RT Ori, V Pav, RT Pav,
TT Per, BU Per, SY Vel.  

\medskip

\noindent
{\bf 4. Discussion}

\smallskip

The stars in Table 1 have LSP SAs which tend to be small; for
almost half, they are 0.1 mag or less.  This is consistent with previous
results in which the SAs are 0.5 mag or less.  Three stars stand out,
however: V Hya, with visual SA = 1.11, BU Lac, with visual SA = 0.88,
and SV And with visual SA = 1.98.
V Hya is a famous variable in which the deep minima are believed to be
due to obscuration by a cloud associated with a companion passing in
front of the red giant -- an extreme example of the process proposed
by Soszy\'{n}ski {\ et al.} (2021).  SV And also has deep minima; my data
are extensive enough to show that the amplitude of the LSP does not
vary significantly over 100 years.  It would be interesting to see whether
the SA correlated with any other properties of the star.

The ratios LSP/P peak strongly at 9-10, which suggests that most of the stars pulsate
in the first overtone.  This is to be expected, since most of the stars
in the table are shorter-period pulsators, rather than large-amplitude
Mira stars which tend to pulsate in the fundamental mode.

In 33 of the 37 stars in Table 1, my LSP is consistent with Houk's.
The ones that disagree are SV And, S Dra, RY Lac, and TW Oph.
 
There are a few stars for which I could find LSPs, but not pulsation
periods.  According to the Soszy\'{n}ski {\it et al.} (2021) hypothesis,
there is no reason why a non-pulsating star could not have an LSP, though
there would have to be some mechanism by which the companion could become
dust-shrouded.

Table 2 lists stars which had adequate data, but in which I was not
able to identify an LSP.  For some of them, the upper limit to the SA was
well below 0.1 mag.  For others, the upper limit was sufficiently high
that I cannot rule out an LSP.

Table 3 lists bimodal stars, and their pulsation period ratios.  These
cluster around 0.5, and are consistent with previous observed (Percy
and Huang 2015) and theoretical values.  There is more scatter than
found by Percy and Huang (2015), but the trend of decreasing P(short)/P(long) with
increasing P(long) is present in our data,

One star had a second period which did not fit either the LSP or the
bimodal classification: RW Psc with periods of 65 and 154 days.  The
former could be a second overtone and the latter a fundamental mode.

Note that Houk (1963) did not include possible bimodal pulsators
in her list of stars -- with one exception: RV Cyg, for which she proposed
periods of 300 and 470 days.  I was not able to obtain results on this star.

\medskip

\noindent
{\bf 5. Conclusions}

\smallskip

Of the 103 red giants listed by Houk (1963) as possibly having LSPs, I
have found evidence for an LSP in 37 of them.  In 33, the LSP that I
have found is consistent with the value given by Houk (1963).  For a few
other stars, I have put stringent upper limits on the the amplitude of
any LSP.  For the rest, I do not have results, or cannot rule out an LSP.
I conclude that the photographic photometrists of the past were rather
successful in detecting and measuring LSPs in pulsating red giants, even
when the amplitudes were small.

There are still many unanswered questions about this phenomenon, which
occurs in a large fraction of RGs.  The Soszy\'{n}ski {\it et al.} (2021)
mechanism for producing LSPs is a fascinating and important one, and
suggests that studies of LSPs should continue.  A combination of long-term
photographic and visual photometry, along with more modern techniques has
the potential to answer some of the unanswered questions.
  
\medskip

\noindent
{\bf Acknowledgements}

\smallskip

This paper is based primarily on data from the AAVSO International
Database.  I am grateful to the observers who made the observations,
and the AAVSO staff who archived them.  I also thank those who created
and who have maintained the VSTAR time-series analysis package which
was used throughout this paper.
I also made use of ASAS-SN photometric data; I thank the
ASAS-SN project team for this remarkable contribution to
stellar astronomy, and for making the data freely available on-line.
The Dunlap
Institute is funded through an endowment established by the David Dunlap
Family and the University of Toronto.

\bigskip

\noindent
{\bf References}

\smallskip

\noindent
Benn, D. 2013, VSTAR data analysis software (http://www.aavso.org/node/803)

\smallskip

\noindent
Houk, N. 1963, {\it Astron. J.}, {\bf 68}, 253.

\smallskip

\noindent
Jayasinghe, T. {\it et al.} 2018, {\it Mon. Not. Roy. Astron. Soc.,} {\bf 477}, 3145. 

\smallskip

\noindent
Jayasinghe, T. {\it et al.} 2019, {\it Mon. Not. Roy. Astron. Soc.}, {\bf 486}, 1907.

\smallskip

\noindent
O'Connell, D.J.K. 1933, {\it Harvard College Observatory Bulletin}, {\bf 893}, 19.

\smallskip

\noindent
Payne-Gaposchkin, C. 1954, {\it Ann. Harvard College Observatory}, {\bf 113}, 189.

\smallskip

\noindent
Percy, J.R. and Huang, D.J. 2015, {\it J. Amer. Assoc. Var. Star Obs.}, {\bf 43}, 118.

\smallskip

\noindent
Percy, J.R. and Deibert, E. 2016, {\it J. Amer. Assoc. Var. Star Obs.} {\bf 44}, 94. 

\smallskip

\noindent
Percy, J.R. and Leung, H. W.-H. 2017, {\it J. Amer. Assoc. Var. Star Obs.}, {\bf 45}, 30. 

\smallskip

\noindent
Percy, J.R. and Fenaux, L. 2019, {\it J. Amer. Assoc. Var. Star Obs.}, {\bf 47}, 202.

\smallskip

\noindent
Samus, N.N. {\it et al.} 2017, {\it Astron. Reports}, {\bf 61}, 80.

\smallskip

\noindent
Shappee, B.J. {\it et al.} 2014, {\it Astrophys. J.}, {\bf 788}, 48. 

\smallskip

\noindent
Soszy\'{n}ski, I., {\it et al}. 2021, {\it Astrophys.J. Lett.}, {\bf 911}, L22.

\medskip

\begin{table}
\caption{Stars with long secondary periods.  The columns list: the star name,
the periods given by Houk (1963), the pulsation period P, the LSP, the LSP SA, and LSP/P -- the last four determined in the present paper.}
\begin{center}
\begin{tabular}{cccccc}
\hline
Star & Houk Periods (d) & P (d) & LSP (d) & SA(LSP) & LSP/P \\
\hline
SV And & 315.96, 930 & 314 & 2283 & 0.84 & 7.3 \\
TW Aur & 150:, 1370 & 144 & 1353 & 0.18 & 9.4 \\
U Cam & 400:, 3000 & 215 & 3000$\pm$200 & 0.13 & 14.0 \\
RS Cam & 88.6, 960 & 90.5 & 999 0.15 & 11.0 \\
RS Cnc & 120, 1700 & 239 & 2050$\pm$100 & 0.07 & 8.6 \\
RT Cnc & 90$\pm$, 542 & 93$\pm$3 & 650$\pm$100 & 0.06 & 7.0 \\
Y CVn & 158, 2000-2200 & 160 & 2009 & 0.06 & 12.6 \\
TZ Car & 69, 1183 & 70: & 1213 & 0.25 & 17.3 \\
VY Cas & 100, 600 & 93.4 & 800$\pm$200 & 0.09 & 8.6 \\
AA Cas & 115, 850 & 75$\pm$5 & 870 & 0.05 & 11.6 \\
SS Cep & 90, 950 & 101.1 & 955 & 0.11 & 9.4 \\
RS CrB & 69.5, 331 & 70 & 332 & 0.24 & 4.7 \\
AF Cyg & 94.1, 960 & 93 & 913: & 0.08 & 9.8 \\
AW Cyg & 220, 2200 & 205: & 2290: & 0.08 & 11.2 \\
S Dra & 136, 1319 & 181.7 & 1000 & 0.08 & 5.5 \\
TX Dra & 78, 654 & 77 & 700$\pm$15 & 0.14 & 9.1 \\
Z Eri & 80, 746 & 78 & 710$\pm$20 & 0.10 & 9.1 \\
X Her & 95.0, 746 & 100 & 700 & 0.08 & 7.0 \\
30 g Her & 70$\pm$, 900 & 88 & 875 & 0.16 & 9.9 \\
V Hya & 533, 6500 & 531 & 6500$\pm$500 & 1.11 & 12.2 \\
RV Lac & 67, 550-700 & 70 & 625$\pm$8 & 0.32 & 8.9 \\
RY Lac & 38-91, 300-420 & 80 & 700$\pm$100 & 0.05 & 8.8 \\
BU Lac & 200, 2000 & 250: & 2200$\pm$200 & 0.88 & 8.8 \\
S Lep & 90, 875-900 & 90: & 867$\pm$10 & 0.23 & 9.6 \\
T Mus & 93, 1021 & 93: & 1020 & 0.23 & 11.6 \\
TW Oph & 165:, 2000 & 376.1 & 3225 & 0.23 & 8.6 \\
W Ori & 212, 2450 & 211 & 2400$\pm$50 & 0.20 & 11.4 \\
RX Peg & 110, 629 & 110 & 650 & 0.12 & 5.9 \\
TW Peg & 90, 956.4 & 100 & 952$\pm$4 & 0.11 & 9.5 \\
rho Per & 33-55, 1100 & 54.65 & 1000$\pm$300 & 0.04 & 18.3 \\
T Per & 326, 2800 & 340 & 2500$\pm$500 & 0.06 & 7.4 \\
U Per & 320.63, 2500 & 320 & 2500$\pm$100 & 0.31 & 7.8 \\
UZ Per & 91, 927 & 90 & 895 & 0.29 & 9.9 \\
RT Psc & 70, 533 & 70: & 515$\pm$5 & 0.08 & 7.4 \\
Y Tau & 240.9, 1750 & 245 & 1733 & 0.04 & 7.1 \\
ST UMa & 81, 590 & 90 & 610$\pm$10 & 0.06 & 6.8 \\
V UMi & 72.0, 760 & 72 & 750$\pm$20 & 0.08 & 10.4 \\
\hline
\end{tabular}
\end{center}
\end{table}

\begin{table}
\caption{Stars possibly lacking long secondary periods.  The columns list: the star
name, the periods given by Houk (1963), the pulsation period P, in days, and
an estimate of the maximum semi-amplitude of the LSP -- the last two determined
in the present paper.}
\begin{center}
\begin{tabular}{cccc}
\hline
Star & Houk Periods & P (d) & A(max) \\
\hline
SS And & 152, 650 & 152: & 0.15 \\
TZ And & 118, 974 & 114.7 & ...\\
UX And & 400$\pm$, 7000 & 402 & 0.10 \\
V Aql & 353, 2270 & 362 & 0.05 \\
UU Aur & 235, 3500 & 445 & 0.07 \\
RX Boo & 78:, 500 & 160 & 0.07 \\
ST Cam & 195, 2100 & 201 & 0.09 \\
RS Cap & 340, 3360 & 130 & 0.15 \\
WZ Cas & 186.0, 4000 & 192 & 0.04 \\
CQ Cas & 110-250, 2300 & 204, 367 & 0.15 \\
CV Cep & 60, 2000 & 60: & 0.18 \\
RR CrB & 60.8, 377 & 55 & 0.04 \\
AW Cyg & 220, 2200 & 204 & 0.04 \\
CH Cyg & 97, 4700 & --- & 0.30 \\
V Eri & 97, 1209 & 300 & 0.30 \\
Y Hya & 95, 302.8 & 363 & 0.10 \\
KP Lyr & 146, 1300 & 140 & 0.13 \\
Y Mic & 182, 4650 & 181 & 0.20 \\
SX Mon & 100$\pm$, 1100 & 100: & 0.10 \\
W Nor & 134.7, 1300 & 147 & 0.20 \\
BQ Ori & 110, 795 & 247 & 0.08 \\
Y Per & 252.3, 2400 & 253 & 0.03 \\
BU Per & 365, 2950 & --- & 0.10 \\
Z UMa & 198, 1560 & 190 & 0.04 \\
TZ Vir & 134, 6900 & 168 & 0.04 \\
\hline
\end{tabular}
\end{center}
\end{table}

\begin{table}
\caption{Stars which are bimodal.  The columns list: the star name, the
longer pulsation period Pa in days, the shorter pulsation period Pb in days,
and the ratio Pb/Pa.}
\begin{center}
\begin{tabular}{cccc}
\hline
Star & Pa (d) & Pb (d) & Pb/Pa \\
\hline
UX And & 402 & 218 & 0.54 \\
UU Aur & 445 & 235 & 0.53 \\
RS Cam & 159 & 89 & 0.56 \\
U Cam & 400 & 215 & 0.54 \\
WZ Cas & 370 & 192: & 0.52: \\
CQ Cas & 367 & 204 & 0.56 \\
SS Cep & 179 & 100 & 0.56 \\
AF Cyg & 174 & 93 & 0.53 \\
RS Dra & 276 & 142 & 0.51 \\
TX Dra & 134 & 77 & 0.57 \\
RY Lac & 80 & 46.5 & 0.58 \\
BQ Ori & 247 & 120: & 0.49: \\
TW Oph & 376 & 172 & 0.46 \\
W Ori & 317 & 211 & 0.67 \\
RW Psc & 154 & 65 & 0.42 \\
V UMi & 123 & 72 & 0.59 \\
\hline
\end{tabular}
\end{center}
\end{table}

\begin{table}
\caption{Stars with missing or inaccurate GCVS periods.  The columns list:
the star name, the GCVS period in days, and the suggested period, based on
the present analysis.}
\begin{center}
\begin{tabular}{ccc}
\hline
Star & GCVS period (d) & suggested period (d) \\
\hline
TZ And & none & 114.7 \\
TW Aur & 104 & 144 \\
RS Cnc & 229.155 & 239 \\
Y CVn & 3000 & 160 \\
TZ Car & none & 70 \\
VY Cas & 116 & 93.4 \\
AA Cas & none & 75 \\
SS Cep & 340 & 100 \\
CV Cep & none & 60 \\
V539 Cyg & 160 & not 160 \\
X Her & 95.0 & 101.5 \\
RY Lac & none & 80 \\
BQ Ori & 110 & 247 \\
ST UMa & 110 & 90 \\
TZ Vir & 134 & 168 \\
\hline
\end{tabular}
\end{center}
\end{table}

\clearpage

\medskip

\noindent
{\bf Appendix: Notes on Individual Stars}

\medskip

The following are notes on individual stars.  The symbol v indicates
results from AAVSO visual observations; V indicates results from AAVSO
photoelectric observations; ASAS-SN indicates results from the All-Sky
Automated Survey for Supernovae; PD and PL indicate results from
Percy and Deibert (2016) and Percy and Leung (2017), respectively; GCVS indicates a period
from the {\it General Catalogue of Variable Stars} (Samus {\it et al}. 2017); numbers
in parentheses are the semi-amplitudes of the periods given; and a colon
denotes uncertainty.

\smallskip

{\it SV And}: Houk: 316d, 930d.  v: 314.96d (1.98); no LSP present. ASAS-SN: 313.0d. 

{\it TZ And}: Houk: 118d, 974d.  v: 114.7d (0.07); no LSP. ASAS-SN: 120:d plus very slow variation; no evidence for the 974d period.  PD: 114.8d, 1355.1d.

{\it UX And}: Houk: 400$\pm$d, 7000d.  v: 219.5d (0.19), 7000-9000d (0.16).  V: 215.8d (0.19), 401.8d (0.36).  ASAS-SN: $\sim$370d.  Bimodal?

{\it V Aql}: Houk: 353d, 2270d.  v: 215.9d (0.06), 362.4d (0.09), LSP not visible.  V: several possible LSPs; none stands out.

{\it TW Aur}: Houk: 150:d, 1370d.  v: 1353d (0.17).  V: 143.6d (0.21).  ASAS-SN  suggests $\sim$120d, $\sim$600d.  104d GCVS period not supported.

{\it UU Aur}: Houk: 235d, 3500d.  v: 235.1d (0.08), 444.6d (0.13), no LSP visible.

{\it RX Boo}: Houk: 78:d, 500d.  v: 160.4d (0.07), LSP not clear.  PD: 160.3d, 2205.1d.

{\it U Cam}: Houk: 400:d, 3000d.  v: 219.5d (0.09), 2941d (0.12).  V: 212.8d (0.13), 2631d (0.23).  ASAS-SN: 400d.  PD: 219.4d, 2967.4d. Possibly bimodal.

{\it RS Cam}: Houk: 88.6d, 960d.  v: 89.4d (0.19), 159.4d (0.10), 1019d (0.14).  ASAS-SN: $\sim$90d.  PD: 90.5d, 999d. Possibly bimodal.

{\it ST Cam}: Houk: 195d, 2100d.  v: 201d (0.09), 363d (0.10), 1557d (0.09).  ASAS-SN: 170d.  Possibly bimodal.

{\it RS Cnc}: Houk: 120d, 1700d.  v: 238.9d (0.13).  V: 239.5d (0.21).  PD: 240.8d, 2050d.  

{\it RT Cnc}: Houk: ~90d, 542d.  v: 93$\pm$3d (0.06), 750$\pm$100d (0.06). V: 93$\pm$3d (0.17), 600$\pm$50d (weak).  ASAS-SN: $\sim$100d.  PD: 89.3d, 691.7d.  Evidence for LSP is weak.

{\it Y CVn}: Houk: 158d, 2000-2200d.  v: 159.2d (0.05); there is an indistinct peak at 2008d.  PD: 160d, 2008.9d.

{\it TZ Car}: Houk: 69d, 1183d.  v: 1213d (0.24), shorter periods are noisy.  ASAS-SN: $\sim$70d.

{\it BZ Car}: Houk: 97d, 1800d.  v: 110.7d (0.09).  ASAS-SN: $\sim$120d.

{\it VY Cas}: Houk: 100d, 600d.  v: 93.4d (0.12), 1014d (0.15).  ASAS-SN: $\sim$100d, $\sim$600d.  LSP uncertain.  116d GCVS period probably needs adjusting.

{\it WZ Cas}: Houk: 186.0d, 4000d.   v: 192.1d (0.06), 369.7d (0.15); LSP amplitude less than 0.04.  V: 370d only.  ASAS-SN: 200d.  Possibly bimodal. 

{\it AA Cas}: Houk: 70-115d, 850d.  v: 74.3d (0.04), 870d (0.04).  ASAS-SN: $\sim$75d.  PD: 80.1d, 866.8d.  LSP probably present.  No GCVS period.

{\it CQ Cas}: Houk: 110-250d, 2300d.  v: 204.8d (0.33), 367.4d (0.40), not aliases.  V: 204d (0.59).  Bimodal, no strong evidence for LSP.

{\it SS Cep}: Houk: 90d, 950d.  v: 100.3d (0.04), 951.5d (0.10).  V: 99.4d (0.10), 179d (0.12), 922.5d (0.33).  ASAS-SN: $\sim$100d, $\sim$1000d.  PD: 101.1d, 955.2d. Possibly bimodal.

{\it CV Cep}: Houk: 60d, 2000d.  V: 731d (0.23).  ASAS-SN: $\sim$60d and 1000+d.

{\it RR CrB}: Houk: 60.8d, 377d.  v: 55.3d (0.04), 91.0d (0.03), 392.8d (0.04).  ASAS-SN consistent with $\sim$60d and $\sim$400d.  PL: 55.5d, 630d.  LSP probably present.  Low amplitudes!

{\it RS CrB}: Houk: 69.5d, 331d.  v: 330.0d (0.24).  ASAS-SN: $\sim$70d, $\sim$300d. 332.2d GCVS period is LSP.

{\it T Crt}: Houk: 70$\pm$, 1006.  ASAS-SN: $\sim$70d.

{\it AF Cyg}: Houk: 94.1d, 960d.  v: 93.4$\pm$1d (0.10), 174$\pm$2d (0.08), 913d or 1439d (0.07). V: 92.8d (0.22), 174$\pm$2d (0.34), 92.8d (0.22).  ASAS-SN: $\sim$160d.  PL: 94d, 1439d.  Probably bimodal with probable LSP.

{\it AW Cyg}: Houk: 220d, 2200d.  v: 357d (0.06); no sign of a 220d period or LSP.  V: 203.8d (0.12), LSP, if present, is weak.  ASAS-SN: $\sim$190d.  PD: 209:d, 2289:d.

{\it CH Cyg:} Houk: 97, 4700.  The light curve of this famous variable is
dominated by large irregular eruptions.

{\it V539 Cyg}: Houk: 160d, 1500d.  v: 355$\pm$10d (0.11).  ASAS-SN: $\sim$90d.  160d GCVS period appears not correct.

{\it S Dra}: Houk: 136d, 1319d.  v: 181.7d (0.07), 311.7d (0.07), 997d (0.08).  ASAS-SN: $\sim$200d.  The Houk and GCVS period of 136d does not seem to be present.

{\it RS Dra}: Houk: 141.15d, 280d.  v: 142$\pm$1d (0.27), 276$\pm$2d; the first may be a harmonic.  ASAS-SN: $\sim$40d and ~300d.  Difficult to interpret; if the 40d ASAS-SN period is real, then one of the longer periods may be an LSP.

{\it TX Dra}: Houk: 78d, 654d.  v: 77.0d (0.08), 136$\pm$1d (0.07), 712.8d (0.14).  V: 77.5$\pm$1d (0.14), 132$\pm$2d (0.24), 868d ((0.17).  ASAS-SN: ~75d.  PD: 77.5d, 711.8d.  Bimodal pulsator with LSP.

{\it UU Dra}: Houk: 120d, 960d.  ASAS-SN: $\sim$120d.  No sign of LSP.

{\it Z Eri}: Houk: 80d, 746d.  v: 78.4d (0.05), 724.6d (0.10).  V: 78.5d (0.11), 730.5d (0.13).  PD: 78.5d, 692.4d.  PL: 118.4d, 725d.  Longer period is an LSP.  The 80d GCVS period needs refining.

{\it X Her}: Houk: 95.0d, 746d.  v: 101.4d (0.06), 700$\pm$30d (0.08).  V: 101.5d (0.19), 700$\pm$30d (0.23).  ASAS-SN: $\sim$100d; an 800d period may be present.  PD: 176.6d, 1185.3d.  PL: 176.6d, 667d.  

{\it 30 g Her}: Houk: 70$\pm$d, 900d.  v: 89d (0.03), 880d (0.16).  V: 88d (0.07), 873d (0.16).   PD: 87.6d, 877d.  LSP is present.

{\it V Hya}: Houk: 533d, 6500d.  v: 531.9d (0.53), 6329d (1.13).  PD: 531.4d, 6907.4d.

{\it Y Hya}: Houk: 95d, 302.8d, 1200d.  v: 363.0d (0.13), no conspicuous LSP.

{\it SU Hya}: Houk: 95$\pm$d, 780d.  ASAS-SN: ~90d, ~500d.

{\it U Lac}: Houk: ~150d, 550-690d.  AAVSO data sparse.  ASAS-SN: $\sim$50d, ~700d (dominant).

{\it RV Lac}: Houk: 67d, 550-700d.  v: 632.9d (0.32).  V: 70d (0.19), 98d (0.19), 617.3d (0.40).  ASAS-SN: $\sim$60d, ~600d.  PD: 70d, 632d.  LSP present. 

{\it RY Lac}: Houk: 38-91d, 300-420d.  v: 46d (0.03), 81d (0.04), 612d (0.05).  V: 47d (0.10), 80d (0.10), 600-800d (0.13).  ASAS-SN: 79.6d, 300-400d.  Bimodal; LSP is not clear. 

{\it BU Lac}: Houk: 200d, 2000d.  GCVS: 2000d.  v: 2128d (0.95).  V: 2403d (1.25); no evidence in v or V or ASAS-SN for 200d period.

{\it S Lep}: Houk: 90d, 875-890d.  v: 857d (0.23), very weak peak at 99d (amplitude less than 0.05); V: 878d (0.31), no peak near 90d. Rather bright for ASAS-SN.

{\it KP Lyr}: Houk: 146d, 1300d.  v: 140.4d (0.21) plus several possible LSPs.  ASAS-SN: ~130d and possibly $\sim$1000d.  No LSP results.

{\it Y Mic}: Houk: 182d, 4650d.  v: 181$\pm$2d (0.17), no clear LSP. 

{\it RV Mon}: Houk: 131.5d, 1047d.  v: 360.0d (0.13), 982.3d (0.12).  LSP is weak.

{\it T Mus}: Houk: 93d, 1021d.  v: 1022.5d (0.23); 93d period not present in our data.

{\it W Nor}: Houk: 134.7d, 1300d.  v: 147.2d (0.19); possible 823d or 2020d LSP.  ASAS-SN is consistent with this pulsation period and an LSP $\ge$ 2000d.

{\it TW Oph}: Houk: 165:d, 2000d.  v: 171.9d (0.16), 376.1d (0.41).  GCVS: 185d.  Bimodal.

{\it W Ori}: Houk: 212d, 2450d.  v: 210.7d (0.12), 316.2d (0.11), 2358d (0.20).  V: 317.3d (0.25).  PD: 210.7d, 2335.1d.

{\it BQ Ori}: Houk: 110d, 795d.  v: 127.3d (0.10), 246.9d (0.17), possibly 709d.  V: 118d (0.25), 248d (0.50).  ASAS-SN: $\sim$100-250d, irregular.  PD: 246.6d, 2147.8d. Bimodal?

{\it V574 Oph}: Houk: 71.5d, 500d.  ASAS-SN: ~70d; possible LSP.

{\it RX Peg}: Houk: 110d, 629d.  V: $\sim$650d, shorter period possibly present. ASAS-SN: $\sim$110d, possible LSP $\ge$ 1500d.

{\it TW Peg}: Houk: 90d, 956.4d.  v: 100.0d (0.04) weak, 948.8d (0.11).  ASAS-SN: $\sim$100d, $\sim$750d.

{\it T Per}: Houk: 326d, 2800d.  v: 2488d (0.06), 326d not present.  V: 342.9d (0.18), 1938d (0.17) uncertain.  ASAS-SN: 318.7d, possible 2000-3000d.  GCVS: 2430d.

{\it Y Per}: Houk: 252.3d, 2400d.  v: 252.8d (0.41), LSP amplitude less than 0.04.  V: 253.8d (0.37), possible LSP.  ASAS-SN: $\sim$200d.

{\it U Per}: Houk: 320.63d, 2500d.  v: 318.57d (1.07), 2531.6d (0.31).  V: 2500$\pm$10d.  ASAS-SN: 320.3d, slow (could be 2500d).

{\it TT Per}: Houk: 82d, 843d.  v: nothing stands out.  ASAS-SN: $\sim$90d.

{\it UZ Per}: Houk: 91d, 927d.  v: 896.9d (0.29), no sign of 91d.  V: no sign of 91d.  ASAS-SN: 90d, $\sim$1000d.  PD: 186.3d, 893.4d.

{\it rho Per}: Houk: 33-55d, 1100d.  V: 54.65d (0.05), 1169d (0.04).  PL: 55d, 723d.

{\it RT Psc}: Houk: 70d, 533d.  v: 511.2d (0.19), no sign of shorter period.  V: 517.6d (0.27), shorter period signal is noisy.

{\it RW Psc}: Houk: 31-53d, 154d.  v: 154.2d (0.15).  ASAS-SN: 65$\pm$15d, ~150d.  Unusual period ratio.

{\it Y Tau}: Houk: 240.9d, 1750d.  v: 244.7d (0.10), no LSP.  V: 1733d (0.36).  ASAS-SN: $\sim$250d.

{\it Z UMa}: Houk: 198d, 1560d.  v: 192$\pm$3d (0.38), no LSP.  V: 188.5d (0.92), 4184d (0.30).  ASAS-SN: $\sim$200d, no evidence for 1560d period.

{\it ST UMa}: Houk: 81d, 590d.  v: 90.2d (0.04), 610.5d (0.06).  V: 89.5d (0.13), ~600d (0.14).  ASAS-SN: 80$\pm$10d, $\sim$600d possibly present.  PD: 90.3d, 623.1d.

{\it V UMi}: Houk: 72.0d, 760d.  v: 72.88d (0.10), 121.1d (0.04), 754,7d (0.08).  V: 71.2d (0.16), 124.8d (0.15), 730d (0.22).  ASAS-SN: $\sim$75d.  PD: 72.9d, 757.3d.  Bimodal.

{\it TZ Vir}: Houk: 134d, 6900d.  v: 167.7d (0.24), 8928d (0.25).  ASAS-SN: consistent with 160d period.

\end{document}